\documentclass[aps,pra,twocolumn,english,showpacs]{revtex4-1}
\usepackage{amsthm}
\usepackage{amsmath}
\usepackage{amsfonts,dsfont,upgreek}
\usepackage{amssymb,epsfig,setspace}
\usepackage{graphicx}
\usepackage{babel}
\usepackage{verbatim}
\newcommand{\Res}{\operatorname{Res}}

\begin{document}

\hyphenation{ano-ther ge-ne-ra-te dif-fe-rent know-le-d-ge po-ly-no-mi-al}
\hyphenation{me-di-um  or-tho-go-nal as-su-ming pri-mi-ti-ve pe-ri-o-di-ci-ty}
\hyphenation{mul-ti-p-le-sca-t-te-ri-ng i-te-ra-ti-ng e-q-ua-ti-on}
\hyphenation{wa-ves di-men-si-o-nal ge-ne-ral the-o-ry sca-t-te-ri-ng}
\hyphenation{di-f-fe-r-ent tra-je-c-to-ries e-le-c-tro-ma-g-ne-tic pho-to-nic}
\hyphenation{Ray-le-i-gh di-n-ger Ham-bur-ger Ad-di-ti-o-nal-ly}
\hyphenation{Kon-ver-genz-the-o-rie ori-gi-nal in-vi-si-b-le cha-rac-te-ri-zed}
\hyphenation{sa-ti-s-fy}
\hyphenation{Ne-ver-the-less sa-tu-ra-te E-ner-gy}

\title{On beautiful analytic structure of the S-matrix}

\author{
Alexander Moroz$^1$ and Andrey E. Miroshnichenko$^2$} 

\affiliation{$^1$Wave-scattering.com\\
$^2$School of Engineering and Information Technology,
University of New South Wales Canberra
Northcott Drive, Campbell, ACT 2600, Australia}

\begin{abstract}
For an exponentially decaying potential, analytic structure of the 
$s$-wave S-matrix can be determined up to the slightest detail, including position of all 
its poles and their residues. Beautiful hidden structures can be revealed by its domain coloring.
A fundamental property of the S-matrix is that any bound state corresponds 
to a pole of the S-matrix on the physical sheet of the complex energy plane.
For a repulsive exponentially decaying potential, none of infinite number of
poles of the $s$-wave S-matrix on the physical sheet corresponds to any physical state.
On the second sheet of the complex energy plane, the S-matrix has infinite number
of poles corresponding to virtual states and a finite number of poles
corresponding to complementary pairs of resonances and anti-resonances.
The origin of redundant poles and zeros is confirmed to be related to peculiarities of 
analytic continuation of a parameter of two linearly independent analytic functions. 
The overall contribution of redundant poles to the asymptotic 
completeness relation, provided that the residue theorem can be applied, 
is determined to be an oscillating function. 
\end{abstract}


\maketitle

%% tar -cvf - rdnt_a.tex rdntbibnjp.tex Fig-S5rep.eps Fig-S5att.eps | gzip > rdntrep.tar.gz

\section{Introduction}
\label{sc:intr}
%%%%%%%%%%%%%%%%%%%%%%%%%%%%%%%%%%%%%
There have been known for long time many exactly solvable models \cite{IH}, yet it has remained rare
that one can determine complete analytic structure of the S-matrix.
As a consequence, the origin of singularities of the S-matrix has not been fully understood.
The latter holds in particular for potentials of infinite range.
Unlike the case of a finite-range potential, the S-matrix for an infinite range potential 
can have {\em infinite} number of poles on the imaginary axis
and there are no theorems known to characterize its essential singularity for $k\to\infty$ \cite{RN},
such as the infinite product representation of the
S-matrix for finite range potentials \cite[eq. (16)]{vanK}, \cite[eq. (12.107a)]{RN}, \cite[eq. (42)]{Hu}.
Furthermore, the S-matrix for infinite range potentials can be plagued 
with {\em redundant} poles and zeros \cite{Ma1,Ma2,Ma3}. The existence of latter was established
over 70 years ago by Ma in the case of the S-matrix 
for an attractive exponentially decaying potential of infinite range.
The {\em redundant} poles and zeros do {\em not} correspond to any bound state, 
half-bound state, (anti)resonance, or a virtual state \cite{Ma1,Ma2,Ma3}. 
This bears important consequences for relating analytic properties of 
the S-matrix to physical states. In particular, the presence of redundant poles on the physical sheet 
of the complex energy plane implies that the S-matrix need not satisfy 
a general condition of Heisenberg \cite{Ma2,tH,BPS}. The issue is of fundamental importance for 
the S-matrix theory. Ma's finding inspired and motivated many authors, 
such as in now classical refs. \cite{vanK,tH,BPS,Jst47,Brg,Ps,PZ}.
There is a noticeable recent revival of interest in the complex analytic structure of the S-matrix
related either to (i) the so-called modal expansion of the scattered field 
\cite{GTV,CPS}, or to (ii) non-Hermitian scattering Hamiltonians \cite{SBK}.
The differences with potentials of finite range highlighted above bear important 
consequences when attempting to generalize modal expansion of the scattered field 
for a finite range potential \cite{GTV,CPS} to potentials of infinite range.

Potentials of infinite range are common in physics (e.g. Coulomb and gravitational fields).
In what follows we analyze the $s$-wave S-matrix for the Schr\"odinger equation 
in a textbook example of an exponentially decaying potential,
\begin{equation}
V_+(r)=V_0 {\rm e}^{-r/a},
\label{edrp}
\end{equation}
where $V_0>0$ and $a>0$ are positive constants and $r$ is radial distance. 
An attractive exponentially decaying potential with an applications to deuteron
was studied as early as in \cite[pp. 100-11]{BB}. Repulsive exponentially decaying potentials play also 
their role in physics \cite{HD,WB}. However, any practical relevance of exponentially decaying potentials 
is not so important in the present case. It is rather that the exponentially decaying potential 
has been at the cornerstone of non-relativistic quantum scattering \cite{vanK,tH,BPS,Jst47,Brg,Ps,PZ}, because 
it enables to analytically determine the S-matrix. Thereby it provides a window into the class of potentials
of infinite range. This has been well recognized by the classics of non-relativistic quantum scattering 
theory \cite{vanK,Ma2,Ma3,tH,BPS,Jst47,Brg,Ps,PZ}. 

The potential (\ref{edrp}) is the {\em repulsive} analogue of the attractive 
exponentially decaying potential, $V_-(r)=-V_+(r)$,
studied by Ma and others \cite{Ma1,Ma2,Ma3,BB}. 
Like its attractive cousin, potential (\ref{edrp}) satisfies conditions \cite[eqs. (12.20) and (12.21)]{RN}
sufficient to prove analyticity of the S-matrix merely in a strip Im $ |k|<2/a$ 
centered around the real axis in the complex plane of momentum $k$ \cite[p. 352]{RN}.
Nevertheless, as it will be shown here, the $s$-wave S-matrix can be 
determined analytically in the whole complex $k$ plane, what the classical monograph \cite{RN} 
surprisingly never mentions. Furthermore, analytic structure of 
the S-matrix can be determined up to the finest detail, including position of all 
its poles and their residues. (Note in passing that even if analytic form 
of the S-matrix is known, a complete determination of the position of all 
its poles and their residues is a nontrivial task, as is also the case here.)
Beautiful hidden structures can be revealed by its domain coloring.
The repulsive example turns out to be rather extreme example in that 
the resulting S-matrix (\ref{MaSrf}) will be shown to have infinite number 
of redundant poles on the physical sheet in the complex energy plane without a single bound state.
At the same time, the resulting S-matrix (\ref{MaSrf}) will be shown to have infinite number
of poles corresponding to {\em virtual} states on the second sheet of the complex energy plane.
Unlike the attractive case \cite{AM2h}, there are obviously no bound states present. 
However, one can identify pairs formed by a resonance (Re $k> 0$) 
and anti-resonance (Re $k< 0$) arranged symmetrically with respect to the 
imaginary axis in the complex $k$-plane, each of them
being absent in the attractive case \cite{AM2h}.

The outline is as follows. 
After preliminaries in Sec. \ref{sc:prln} we provide a rigorous canonical analysis of the $s$-wave S-matrix 
along the lines of monograph \cite{RN} in Sec. \ref{sc:nsm}. We obtain analytic expressions 
for the Jost functions and determine the S-matrix (\ref{MaSrf}). This enables us to illustrate 
the validity of general theorems in the slightest detail and to achieve a deep understanding 
of the analytic structure of the S-matrix. An indispensable part of the analysis are
Coulomb's results \cite{Clb} on zeros of the modified Bessel function
$I_\nu(x)$ for fixed nonzero argument $x\in\mathbb{R}$ considered as a complex 
entire function of its order $\nu$, which are summarized in supplementary material.
The origin of redundant poles is analyzed in Sec. \ref{sc:rdntp}.
In the attractive case, $V_-(r)$,
the origin of redundant poles and zeros has recently been related to peculiarities of analytic 
continuation of a parameter of two linearly independent analytic solutions of a second 
order ordinary differential equation \cite{AM2h}. The crux of the appearance of redundant 
poles and zeros lied in that analytic continuation of a parameter of two linearly 
independent solutions resulted in linearly dependent 
solutions at an infinite discrete set of isolated points of the parameter complex plane \cite{AM2h}.
In what follows we confirm the recent observation also in the repulsive case.
In Sec. \ref{sc:hgc} devoted to the Heisenberg condition we analytically determine 
the residues of the S-matrix of redundant poles [eq. (\ref{rprs})] and the overall 
contribution of redundant poles [eq. (\ref{tcrp})]
in the asymptotic completeness relation [eq. (\ref{hgc})], 
provided that the contribution can be evaluated by the residue theorem.
For the sake of completeness, in Sec. \ref{sc:vst} we analytically determine the residues of 
virtual states. We end up with discussion of a number of important issues and conclusions.

\section{Preliminaries}
\label{sc:prln}
%%%%%%%%%%%%%%%%%%%%%%%%%%%%%%%%%%%%%
It is straightforward to modify the essential steps of Bethe and Bacher \cite[p. 108]{BB}
in analyzing the attractive potential $V_-(r)$ to the present case of 
the repulsive potential $V_+(r)$ (\ref{edrp}). With the substitution $\psi(r)=u(r)/r$ one has
\begin{eqnarray}
\psi''(r)+ \frac{2}{r}\, \psi'(r) &=& \frac{u''(r)}{r},
\nonumber
\end{eqnarray}
where prime denotes derivative with respect to the function argument.
Therefore, the radial $s$-wave Schr\"odinger equation for a particle of mass $m$ 
and energy $E$ takes the form
\begin{equation}
u''(r)+[k^2 - U_0 {\rm e}^{-r/a}] u=0,~~ U_0=-\frac{2m}{\hbar^2}\, V_0>0,
\label{edper}
\end{equation}
where $k^2=2mE/\hbar^2$ and $\hbar$ is the reduced Planck constant.
The general solution of (\ref{edper}) is a linear combination of modified 
Bessel functions of the first kind with a complex order ${\rm i}\rho$ 
\cite[Sec. 9:6]{AS}, \cite[Sec. 10.25]{Ol}, where
$\rho=2ak$ is a dimensionless momentum parameter,
\begin{eqnarray}
u(r) &=& c_1 I_{{\rm i}\rho}\left(2a{\rm e}^{-r/(2a)}\sqrt{U_0} \right)
\nonumber\\
&& +\, c_2 K_{{\rm i}\rho}\left(2a{\rm e}^{-r/(2a)} \sqrt{U_0} \right),
\label{gedps}
\end{eqnarray}
with $c_1$ and $c_2$ being arbitrary integration constants.
Indeed, with {\em e.g.} $c_1=1$, $c_2=0$, and $\sigma={\rm e}^{-r/a}$, 
$x=2a\sqrt{U_0\sigma}$, one has
\begin{eqnarray}
u'(r) &=& -\sqrt{U_0 \sigma} I_{{\rm i}\rho}'(x),
\nonumber\\
u''(r) &=& U_0\sigma I_{{\rm i}\rho}''(x) +\frac{\sqrt{U_0\sigma}}{2a}\, I_{{\rm i}\rho}'(x).
\nonumber
\end{eqnarray}
When substituting back into (\ref{edper}), one arrives at 
\begin{equation}
U_0\sigma I_{{\rm i}\rho}''(x) +\frac{\sqrt{U_0\sigma}}{2a}\,
 I_{{\rm i}\rho}'(x) + (k^2-U_0\sigma) I_{{\rm i}\rho}(x)=0.
\nonumber
\end{equation}
After multiplication by $4a^2$,
\begin{equation}
 x^2 I_{{\rm i}\rho}''(x) +x\, I_{{\rm i}\rho}'(x) - [x^2+({\rm i}\rho)^2] I_{{\rm i}\rho}(x)=0,
 \label{dbfe}
\end{equation}
which is the defining equation of the modified Bessel functions of imaginary order 
$\nu={\rm i}\rho$ (cf. \cite[(9.6.1)]{AS}, \cite[(10.25.1)]{Ol}).

\section{A rigorous analysis of the $s$-wave S-matrix}
\label{sc:nsm}
%%%%%%%%%%%%%%%%%%%%%%%%%%%%%%%%%%%%%%%%%%%%%%%%%%%%%%%%%%%

\subsection{The {\em regular} solution $\varphi(r)$}
%%%%%%%%%%%%%%%%%%%%%%%%%%%%%%%%%%%%%%%%%%%%%%%%%%%%
The pair $\{I_{{\rm i}\rho}(z),K_{{\rm i}\rho}(z)\}$ yields always two linearly independent solutions 
of eq. (\ref{dbfe}) and its Wronskian is never zero (cf. \cite[(9.6.15)]{AS}, \cite[(10.28.2)]{Ol}). 
The {\em regular} solution of (\ref{edper}) vanishing at the origin $r=0$ becomes
in the notation of ref. \cite{RN}
\begin{eqnarray}
\varphi(r) &=& C\left[ K_{{\rm i}\rho}(\alpha) I_{{\rm i}\rho}\left(\alpha {\rm e}^{-r/(2a)} \right)\right.
\nonumber\\
&& \left. - I_{{\rm i}\rho}(\alpha) K_{{\rm i}\rho}\left(\alpha {\rm e}^{-r/(2a)} \right)\right],
\label{gedprsi}
\end{eqnarray}
where $\alpha=2a\sqrt{U_0}=\lim_{r\to 0} x(r)$ and 
$C=-2a$ ensures normalization $\varphi'(0)=1$ \cite[eq. (12.2)]{RN}.
Indeeed, for $x\to\alpha$, or equivalently $r\to 0$, 
one finds on using \cite[(9.6.15)]{AS}, \cite[(10.28.2)]{Ol}
\begin{eqnarray}
\varphi'(r) &=& -\frac{\alpha C}{2a} \left[ K_{{\rm i}\rho}(\alpha) 
 I_{{\rm i}\rho}'(x) - I_{{\rm i}\rho}(\alpha) K_{{\rm i}\rho}'(x)\right]
\nonumber\\
&=&- \frac{\alpha C}{2a} W\{ K_{{\rm i}\rho}(x), I_{{\rm i}\rho}(x)\} \to - \frac{C}{2a}\cdot
\nonumber %%\\label{gedprsid}
\end{eqnarray}

\subsection{Irregular solutions $f_\pm(k,r)$}
%%%%%%%%%%%%%%%%%%%%%%%%%%%%%%%%%%%%%%%%%%%%%%%
According to \cite[(9.6.7,9)]{AS}, \cite[(10.30.1-2)]{Ol}, one finds in the limit $z\to 0$,
\begin{eqnarray}
I_\nu(z) &\sim& \frac{z^\nu}{2^\nu \Gamma(\nu+1)}\qquad (\nu\ne -1, -2, -3, \ldots),
\label{as9.6.7}
\\
K_\nu(z) &\sim& \frac12\, \Gamma(\nu) \frac{2^\nu}{z^\nu},\qquad \mbox{Re }\nu >0,
\label{as9.6.9}
\end{eqnarray}
where $\Gamma$ denotes the usual gamma function \cite[sec. 6.1)]{AS}, \cite[sec. 5]{Ol}.
At the same time, for $z\in\mathbb{C}$ \cite[(9.6.6)]{AS}, \cite[(10.27.1-2)]{Ol}:
\begin{equation}
I_{-n}(z)= I_n(z),\qquad K_{-\nu}(z)=K_{\nu}(z).
\label{ipmm}
\end{equation}

Given the asymptotic (\ref{as9.6.7}), (\ref{as9.6.9}), the usual irregular solution for $k\in\mathbb{R}$ 
has to be proportional to $I_{-{\rm i}\rho}(x)$,
\begin{eqnarray}
f_+(k,r)= \Gamma(1-{\rm i}\rho)\left(\frac{\alpha}{2}\right)^{{\rm i}\rho} I_{-{\rm i}\rho}(x).
\label{f+}
\end{eqnarray}
The asymptotic (\ref{as9.6.7}) implies for Im $k\ge 0$ [Re $(-{\rm i}\rho)\ge 0$]
\begin{equation}
f_+(k,r) \sim {\rm e}^{{\rm i}kr},
\nonumber %%\label{Isas}
\end{equation}
showing the characteristic outgoing spherical wave behaviour of $f_+(k,r)$ for $r\to\infty$, $k\in\mathbb{R}$,
and yields $f_+(k,r)$ as exponentially decreasing for $r\to\infty$, Im $k> 0$, in accordance
with general theorems \cite[Sec. 12.1.4]{RN}. 
$f_+(k,r)$ cannot be proportional to $K_{{\rm i}\rho}(x)$, because for ${\rm i}\rho \pi\ne n\pi$, $n\in\mathbb{Z}$, 
\cite[(9.6.2)]{AS}, \cite[(10.27.2)]{Ol}
\begin{equation}
K_{{\rm i}\rho}(x)=
\frac{\pi}{2 \sin ({\rm i}\rho \pi)}\, [I_{-{\rm i}\rho}(x)- I_{{\rm i}\rho}(x)].
\label{imnuknum}
\end{equation}
Then the asymptotic $f_+(k,r)$ would comprise both $\sim {\rm e}^{{\rm i}kr}$ and $\sim {\rm e}^{-{\rm i}kr}$
terms.

Given the analyticity of $\Gamma(1-{\rm i}\rho)$ and $I_{-{\rm i}\rho}$ \cite{Ann}, one can easily verify
$f_+(k,r)$ to be an analytic function
of $k$ regular for Im $k>0$ and continuous with a continuous $k$ derivative in
the region Im $k\ge 0$ for each fixed $r>0$. The second linearly independent irregular solution 
$f_-(k,r)=f_+(k{\rm e}^{{\rm i}\pi},r)\propto I_{{\rm i}\rho}(x)$ 
(assuming an analytic continuation via the upper half plane)
is uniquely determined by the boundary condition $f_-(k,r) \sim {\rm e}^{-{\rm i}kr}$ for $r\to\infty$.

According to \cite[(9.6.2)]{AS}, \cite[(10.27.2)]{Ol}, $I_{-{\rm i}\rho}(z)$ is 
related to $K_{{\rm i}\rho}(z)$ for any ${\rm i}\rho$ by 
\begin{equation}
I_{-{\rm i}\rho}(z)= I_{{\rm i}\rho}(z) + \frac{2}{\pi}\, \sin ({\rm i}\rho \pi)\, K_{{\rm i}\rho}(z).
\nonumber %%\label{imnuknu}
\end{equation}
This enables one to express $f_+(k,r)$ in the basis of $\{I_{{\rm i}\rho}(z),K_{{\rm i}\rho}(z)\}$,
\begin{eqnarray}
\lefteqn{
f_+(k,r)= \Gamma(1-{\rm i}\rho)\left(\frac{\alpha}{2}\right)^{{\rm i}\rho}\times
}
\nonumber\\
 && \left[I_{{\rm i}\rho}(x) + \frac{2}{\pi}\, \sin ({\rm i}\rho \pi)\, K_{{\rm i}\rho}(x)\right].
\nonumber %%\label{f+be}
\end{eqnarray}

\subsection{The Jost functions ${\cal F}_\pm(k)$}
%%%%%%%%%%%%%%%%%%%%%%%%%%%%%%%%%%%%%%%%%%%%%%%%%%%
Using the latter expression, it is straightforward to determine the Jost function \cite[Sec. 12]{RN}
\begin{eqnarray}
\lefteqn{
{\cal F}_+(k):=W_r\{f_+,\varphi\}= C \Gamma(1-{\rm i}\rho)\left(\frac{\alpha}{2}\right)^{{\rm i}\rho}\times
}
\nonumber\\
 && \left[ I_{{\rm i}\rho}(\alpha) 
+\frac{2}{\pi}\, \sin ({\rm i}\rho \pi)\, K_{{\rm i}\rho}(\alpha)\right] W_r\{K_{{\rm i}\rho},I_{{\rm i}\rho}\}
\nonumber\\
 &&
= C I_{-{\rm i}\rho}(\alpha) \Gamma(1-{\rm i}\rho)\left(\frac{\alpha}{2}\right)^{{\rm i}\rho}
\frac{1}{\alpha {\rm e}^{-r/(2a)}} \left(\alpha {\rm e}^{-r/(2a)} \right)'
\nonumber\\
 &&
= I_{-{\rm i}\rho}(\alpha) \Gamma(1-{\rm i}\rho)\left(\frac{\alpha}{2}\right)^{{\rm i}\rho}
\nonumber\\
 &&
= {}_0F_1(1-{\rm i}\rho,\alpha^2/4),
\label{Fkd}
\end{eqnarray}
where the respective $W_r\{.\, ,.\}$ and prime denote the Wronskian 
(cf. \cite[(9.6.15)]{AS}, \cite[(10.28.2)]{Ol}) 
and derivative with respect to $r$, and ${}_0F_1$ is the confluent hypergeometric function.

The complementary Jost function ${\cal F}_-(k):=W_r\{f_-,\varphi\}$ is obtained by replacing 
${\rm i}\rho\to -{\rm i}\rho$ in the above expression for ${\cal F}_+$.
Analytic properties of the Jost functions ${\cal F}_\pm(k)$ follow from that
(i) each branch of $I_{\pm\nu}(z)$ is entire in $\nu$ for fixed $z$ ($\ne 0$) and (ii) $\Gamma(z)$ 
is holomorphic in $z$ having only simple poles at $z=-n$, $n\ge 0$ \cite{AS,Ol}.

\subsection{A detailed analytic structure of the S-matrix}
%%%%%%%%%%%%%%%%%%%%%%%%%%%%%%%%%%%%%%%%%%%%%%%%%%%%%%%%%
The $s$-wave S-matrix is determined as the ratio $S(k)={\cal F}_-(k)/{\cal F}_+(k)$ 
(cf. \cite[eq. (12.71)]{RN})
\begin{equation}
S(k)= \frac{I_{{\rm i}\rho}(\alpha) \Gamma(1+{\rm i}\rho)}{I_{-{\rm i}\rho}(\alpha)\Gamma(1-{\rm i}\rho)}
\left(\frac{\alpha}{2}\right)^{-2{\rm i}\rho},
\label{MaSrf}
\end{equation}
where $k$ dependence here enters through the dimensionless momentum parameter $\rho=2ak$. 
It follows straightforwardly that
%%%%%%%%%%%%%%%
\begin{itemize}
 
\item 
$S(k)$ vanishes either when $I_{{\rm i}\rho}(\alpha)=0$ or, at
the poles of $\Gamma(1-{\rm i}\rho)$ \cite{Gmf}. 

\item
The poles of $S(k)$ are the poles 
of $\Gamma(1+{\rm i}\rho)$ \cite{Gmf} and the zeros of $I_{-{\rm i}\rho}(\alpha)$.

\end{itemize}
%%%%%%%%%%%%
The poles of $\Gamma(1+{\rm i}\rho)$ occur for any ${\rm i}\rho=-n,\, n\in\mathbb{N}_+$,
or $k=k_n$, $k_n={\rm i}n/(2a)$, $n\ge 1$. 
Those are the only poles, and all those poles are simple \cite[(6.1.3)]{AS}.
They give rise to infinite number of simple redundant poles
of $S(k)$ on the positive imaginary axis, {\em i.e.} on the 
physical sheet of the complex energy plane.

The zeros of $I_{-{\rm i}\rho}(\alpha)$ are the only zeros of ${\cal F}_+(k)$, which
is analytic (without any singularity) on the physical sheet (Im $k\ge 0$).
A point of crucial importance for further discussion are the following results on the 
zeros of $I_\nu(x)$, some of which [({\bf a}) and ({\bf c}) below; cf. \ref{Izeros}] 
being well hidden and largely forgotten and, surprisingly, cannot be found either in ultimate 
tables \cite{AS,Ol} or monograph \cite{Wat}:
%%%%%%%%%%%%%%%
\begin{itemize}
 
\item[({\bf a})] $I_\nu(x)$ with fixed nonzero $x\in\mathbb{R}$ and Re $\nu>-3/2$ has no complex zero 
({\em i.e.} with nonzero imaginary part) when considered as a function of its order $\nu$ \cite{Clb}. 
All the roots of $I_\nu(x)$ are {\em real} for Re $\nu>-3/2$.

\item[({\bf b})] $I_\nu(x)$ is real and positive 
for real order $\nu\ge -1$ and $x>0$ \cite[eqs. (9.6.1), (9.6.6)]{AS}.
 
\item[({\bf c})] The roots of $I_\nu(z)$ are asymptotically 
near the negative integers for large $n$ for $0<|z|\ll 1$ and/or $|\nu|\gg |z|+1$ \cite{Clb,Chn}.
Indeed, in the latter asymptotic range one has for the roots of $J_\nu(z)$ \cite[eq. (8)]{Chn}
\begin{equation}
\nu_n\sim -n + \frac{(z/2)^{2n}}{n!(n-1)!}- \frac{(z/2)^{2(n+1)}}{(n+1)!(n-1)(n-1)!},
\label{ch8}
\end{equation}
from which the asymptotic of the roots of $I_\nu(z)$ follows on substituting $z\to iz$.
Obviously, the roots $\nu_n$ of $I_\nu(z)$ are {\em real} for real $x$ when formula (\ref{ch8}) holds.
Worth of noting is that, unlike the roots of $J_\nu(x)$, the roots of $I_\nu(x)$ 
need not be in general simple \cite{Clb}.
 
\end{itemize}
%%%%%%%%%%%%
In the case of $I_{-{\rm i}\rho}(\alpha)$ in the denominator of 
the S-matrix (\ref{MaSrf}), the condition Re $\nu>-3/2$ translates into Im $k>-3/(4a)$.
Therefore, the S-matrix $S(k)$ does not have any singularities for Im $k>-3/(4a)$,
except those on the imaginary axis \cite{Gmf}. The property ({\bf b}) excludes any zero of 
$I_{-{\rm i}\rho}(\alpha)$ on the imaginary axis for Im $\rho\ge -1$ [Im $k\ge -1/(2a)$].
In particular, the property ({\bf b}) prohibits any bound state on the positive imaginary axis.
%%%%%%%%%%
\begin{figure}
\begin{center}
\includegraphics[width=8cm,clip=0,angle=0]{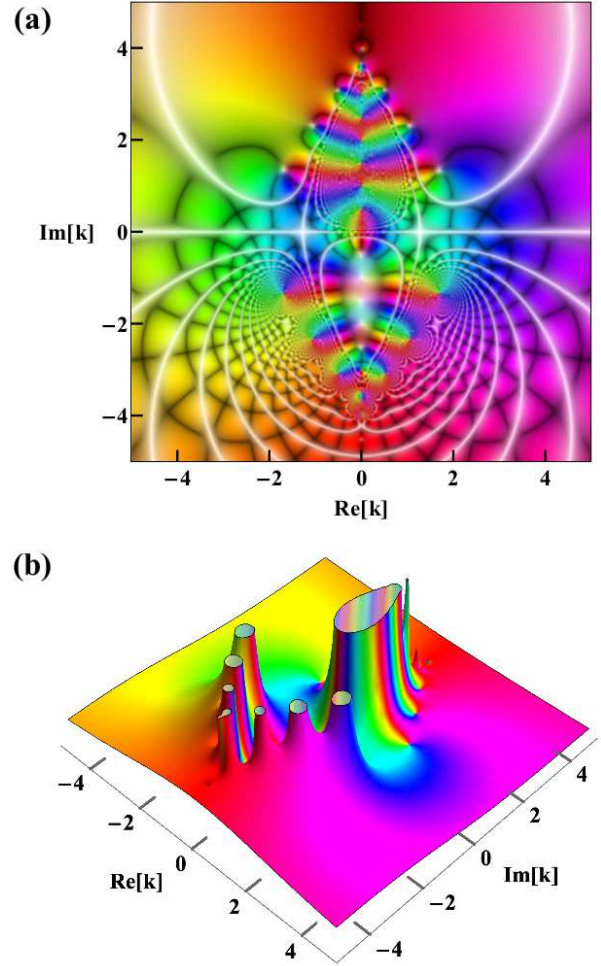}
\end{center}
\caption{Domain coloring (a) and 3D plot (b) of the S-matrix (\ref{MaSrf}) for $\alpha=5$ 
and $a=1$ resembling a crayfish. Hue reflects the phase of the S-matrix. A pole (zero) can be 
identified by a counterclockwise (clockwise) rotated colour wheel (and a white spot) around it.
One can identify redundant poles on the positive imaginary axis at any $k=k_n=in/2$, $n\ge 1$,
poles in the lower half complex plane corresponding to resonance (Re $k> 0$) 
and anti-resonance (Re $k< 0$) pairs arranged symmetrically with respect to the imaginary 
axis at $k\approx \pm 1.690-{\rm i} 1.345,\pm 1.053-{\rm i} 2.314,
\pm 0.509-{\rm i} 3.088$, and virtual states on the negative imaginary axis 
at $k\approx-{\rm i} 3.661, -{\rm i} 3.970, -{\rm i} 4.502$.
}
\label{fgSmt}
\end{figure}
%%%%%%%%%%
Hence, in addition to the redundant poles at $k=k_n$ on the positive imaginary axis, 
the S-matrix (\ref{MaSrf}) can have only poles corresponding to {\em virtual} 
states on the negative imaginary axis in the complex $k$-plane, {\em i.e.} on the 
second sheet of the complex energy plane. The implicit condition on the $k$-values 
of {\em virtual} states is, in virtue of the property ({\bf b}),
\begin{equation}
I_{-2as}(\alpha)=0, \qquad s> 1/(2a),
\nonumber %%\label{BBc}
\end{equation}
where $k=-is$, $s\in\mathbb{R}$. The number of virtual states is {\em infinite}.
For $0<|\alpha|\ll 1$ and/or $2as \gg \alpha+1$, the position
of virtual states can be estimated from (\ref{ch8}) for $x=i\alpha$.
Interestingly, (\ref{ch8}) implies that the roots of $I_{-{\rm i}\rho}(\alpha)$
approach asymptotically the poles of $\Gamma(1-{\rm i}\rho)$ in the denominator
of the S-matrix (\ref{MaSrf}).

%%%%%%%%%%
\begin{figure}
\begin{center}
\includegraphics[width=8cm,clip=0,angle=0]{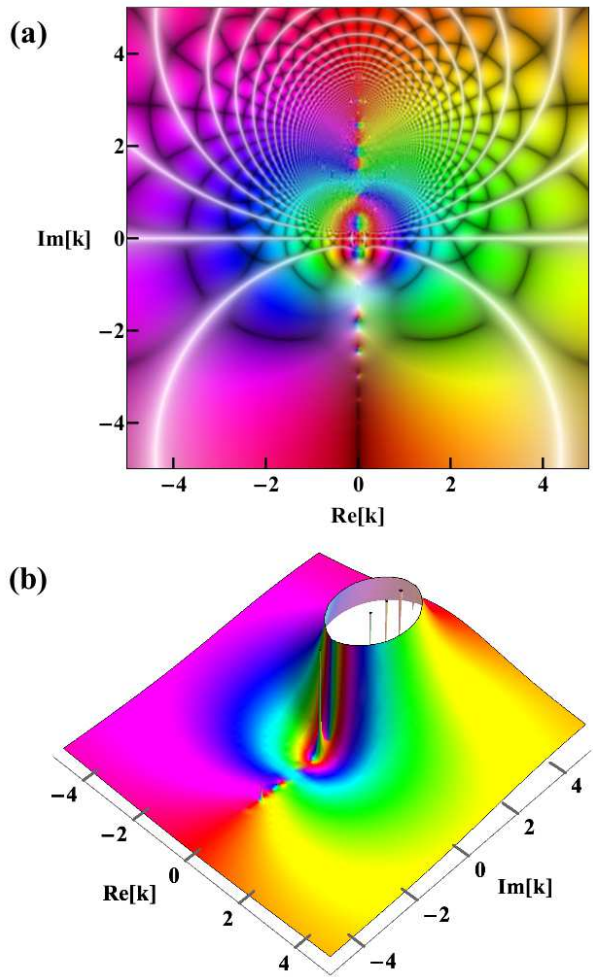}
\end{center}
\caption{Domain coloring (a) and 3D plot (b) of the S-matrix in the attractive case, $V_-(r)$, 
for $\alpha=5$ and $a=1$ resembling a spider in a web.
One can identify redundant poles at any $k=k_n=in/2$, $n\ge 1$, and a single 
bound state at $k\approx {\rm i} 0.947$
on the positive imaginary axis, and poles on the negative imaginary axis 
corresponding to virtual states at $k\approx -{\rm i} 0.163, -{\rm i} 1.032, 
-{\rm i} 1.765, -{\rm i} 2.402, -{\rm i}2.969, -{\rm i}3.4931, 
-{\rm i}3.999,\newline -{\rm i}4.499, -{\rm i} 4.999$.
In contrast to fig. \ref{fgSmt}, any resonance or anti-resonance is missing. 
}
\label{fgSmta}
\end{figure}
%%%%%%%%%%
Last but not the least, for Im $k\le -3/(4a)$ one cannot exclude the 
presence of resonances (Re $k> 0$) and/or anti-resonances (Re $k< 0$) 
in the lower half of the complex $k$-plane outside the imaginary axis.
For $x\in\mathbb{R}$ one has $\overline{I_\nu(x)}=I_{\bar\nu}(x)$ (cf. 
\cite[(9.6.10)]{AS}, \cite[(10.25.2)]{Ol}), which is the Schwarz reflection in $\nu$-variable.
Hence a complex zero $\nu=\nu_0$ implies that also $\bar\nu_0$ is zero, i.e. complex zeros occur 
necessarily in complex conjugate pairs. 
In the complex $k$-plane this translates to that
resonances and anti-resonances form pairs arranged symmetrically with respect to
the imaginary axis. In virtue of the above property ({\bf c}) of the roots $\nu_n$ of $I_\nu(z)$,
one can expect to find complex roots $\nu_n$ only for $|\nu|\sim |z|\not\ll 1$.
This is what is indeed confirmed in fig. \ref{fgSmt}.

The singularities in the complex $k$-plane with negative imaginary part correspond to states 
that do not belong to the Hilbert space since they are not normalizable. However, they
can produce observable effects in the scattering amplitudes, in
particular when they approach the real axis. In the present case, virtual states together
with resonances and anti-resonances cannot approach the real axis in the complex $k$-plane 
closer than the minimal distance $d=3/(4a)$. 

For a comparison, fig. \ref{fgSmta} shows the S-matrix in the attractive case, $V_-(r)$, 
for $a=1$ and $\alpha=5$, {\rm e.g.} for the same parameters as in fig. \ref{fgSmt}.
The S-matrix in the attractive case differs from (\ref{MaSrf}) in that
$I_{\pm{\rm i}\rho}(\alpha)$ are substituted by $J_{\pm {\rm i}\rho}(\alpha)$ \cite{Ma2,AM2h}.
One can identify redundant poles and a single bound state
on the positive imaginary axis, and some of infinite number of virtual states on the negative imaginary axis.
However, any resonance or anti-resonance is forbidden. This is because 
$J_{\nu}(x)$ considered as a function of its order $\nu$ does not have any 
complex pole ({\em i.e.} with nonzero imaginary part) for real $x\ne 0$ \cite{Clb}. All the roots
$\nu_n$ of $J_\nu(x)$ are {\em real} and simple \cite{Clb}.
For large $n$ for $0<|z|\ll 1$ and/or $|\nu|\gg |z|+1$ the roots $\nu_n$ of $J_\nu(z)$ are asymptotically 
near the negative integers according to (\ref{ch8}) \cite{Clb,Chn}.

The present $f_\pm(k,r)$ satisfy all the classical requirements \cite{RN}.
The usual analytic connection between the positive and negative real
$k$ axis, $f_-(k,r)=f_+(k{\rm e}^{{\rm i}\pi},r)$,
together with the boundary condition satisfied by $f_\pm$ leads to
$S(-k) = S^*(k) = S^{-1}(k)$ for any $k\in\mathbb{R}$ \cite[eq. (12.74)]{RN}.
For general $k\in\mathbb{C}$ one has (cf. \cite[eqs. (12.24a), (12.32a)]{RN}, \cite{Hu})
\begin{equation}
S(-k)=S^{-1}(k), \qquad S^*(k^*) = S^{-1}(k).
\label{rn12.32}
\end{equation}
The first relation here can be readily verified for the S-matrix (\ref{MaSrf}).
In order to verify the second of relations (\ref{rn12.32}), 
notice that all the special functions involved in (\ref{MaSrf}) satisfy the 
Schwarz reflection principle $F(\bar{z})=\overline{F(z)}$ in variable $z={\rm i}\rho$ 
for $\alpha\in\mathbb{R}$. Hence
\begin{equation}
S^*(k^*) = \frac{I_{-{\rm i}\rho}(\alpha) 
\Gamma(1-{\rm i}\rho)}{I_{{\rm i}\rho}(\alpha)\Gamma(1+{\rm i}\rho)}
\left(\frac{\alpha}{2}\right)^{2{\rm i}\rho}=S(-k),
\nonumber %%\label{MaSrfcc}
\end{equation}
which is obviously $S^{-1}(k)$.

According to the second of relations (\ref{rn12.32}), to any pole $k_p$ of $S(k)$ on the 
first physical sheet of energy (Im $k> 0$) there corresponds at the position $k_z=k_p^*$,
mirror symmetric with respect to the real axis, a zero
of $S(k)$ on the second sheet (Im $k< 0$), and vice versa \cite[Sec. 12.1.4]{RN}.
According to the combination of first and second of relations (\ref{rn12.32}), 
for any pole $k_p$ of $S(k)$ {\em not} on the imaginary axis there must
be a zero at $k_z=-k_p$, and hence another pole at at $k=-k_p^*\ne k_p$, 
a mirror symmetric position with respect to the imaginary axis. 
The above pole-zero and pole-pole correspondences resulting from (\ref{rn12.32}) 
are nicely reflected in figs. \ref{fgSmt}, \ref{fgSmta}.
Another consequence of symmetry relations (\ref{rn12.32}) is that a {\em simple}
imaginary pole (zero) has to stay on the imaginary axis when model parameters are varied.
Otherwise (\ref{rn12.32}) would be immediately violated.

\section{On the origin of redundant poles}
\label{sc:rdntp}
%%%%%%%%%%%%%%%%%%%%%%%%%%%%%%%%%%%%%%%%%%%
The Wronskian of $I_{\pm {\rm i}\rho}$ \cite[eq. (9.6.14)]{AS}, \cite[eq. (10.28.1)]{Ol},
\begin{equation}
 W_x\{I_{{\rm i}\rho}(x),I_{-{\rm i}\rho}(x)\}=-\frac{2 \sin ({\rm i}\rho \pi)}{\pi x},
\label{wi+-}
\end{equation}
vanishes whenever ${\rm i}\rho\in\mathbb{Z}$.
In the special case ${\rm i}\rho=-n$, $n\in\mathbb{N}$, one finds on 
combining eqs. (\ref{as9.6.7}), (\ref{ipmm}):
\begin{equation}
I_{\pm n}(z)\sim \frac{z^n}{2^n \Gamma(n+1)},\qquad n\in\mathbb{N}.
\nonumber %%\label{as9.6.7m}
\end{equation}
Therefore, the basis $\{I_{{\rm i}\rho}(z),I_{-{\rm i}\rho}(z)\}$ of solutions 
of eq. (\ref{dbfe}) collapses into linearly {\em dependent} 
solutions for any ${\rm i}\rho\in\mathbb{Z}$. 
(For $\rho=0$ the two Bessel functions degenerate into a single one.)
Our recent treatment of attractive potential $V_-(r)$ \cite{AM2h}
suggests that the collapse of the pair $\{I_{{\rm i}\rho}(x),I_{-{\rm i}\rho}(x)\}$ of solutions 
of eq. (\ref{dbfe}) into linearly {\em dependent} solutions 
and the occurrence of redundant poles and zeros at exactly the same points
${\rm i}\rho\in\mathbb{Z}$ is not coincidental.
Like in \cite{AM2h}, it is important to notice that eqs. (\ref{f+}), (\ref{Fkd})
imply {\em factorization} of $f_\pm(k,r)$ as
\begin{equation}
f_\pm(k,r)= \frac{ {\cal F}_\pm (k)}{I_{\mp {\rm i}\rho}(\alpha)}\, I_{\mp {\rm i}\rho}(x),
\label{fctrz}
\end{equation}
where the first factor including the Jost function, ${\cal F}_\pm$, is only a function of $k$, 
and only the second factor, $I_{\mp {\rm i}\rho}(x)$, depends on both $k$ and $r$.
In virtue of (\ref{Fkd}), the first factor is finite for any $I_{\mp {\rm i}\rho}(\alpha)=0$.

Let us ignore for a while the first $k$-dependent prefactors in (\ref{fctrz}).
Then $f_-(k,r)$, which is typically exponentially {\em increasing} on the physical 
sheet as $r\to \infty$, would become suddenly exponentially {\em decreasing} for 
$r\to \infty$ for any ${\rm i}\rho\in\mathbb{Z}_-$, {\em i.e.} $k=k_n$, $n\ge 1$, on the physical sheet, 
very much the same as $f_+(k,r)$. Similarly, $f_+(k,r)$, which is expected to be 
exponentially {\em increasing} on the second sheet for $r\to \infty$,
would become suddenly exponentially {\em decreasing} in the limit 
for any ${\rm i}\rho\in\mathbb{Z}_+$, or $k=-k_n$, $k_n={\rm i}n/(2a)$, $n\ge 1$, on the second sheet, 
very much the same as $f_-(k,r)$.
The role of the $k$-dependent prefactors ${\cal F}_\pm$ is to hide such an 
``embarrassing'' behaviour by causing
the respective irregular solutions $f_\pm(k,r)$ to become {\em singular} at the 
incriminating points ({\em i.e.} $f_-(k,r)$ at $k=k_n$, and $f_+(k,r)$ at $k=-k_n$).
Note in passing that although $f_+(k,r)$ ($f_-(k,r)$) is, for each $r$, an analytic function
of $k$ regular for Im $k>0$ (Im $k<0$) and continuous with a continuous $k$ derivative in
the region Im $k\ge 0$ (Im $k\le 0$), this no longer holds for Im $k<0$ (Im $k>0$). 

The singular prefactors ensure that, in spite of the linear dependency 
of the pair $\{I_{{\rm i}\rho}(x),I_{-{\rm i}\rho}(x)\}$ for any ${\rm i}\rho\to\mathbb{Z}$, the identity 
$W\{f_+,f_-\}=-2{\rm i}k$ \cite[eq. (12.27)]{RN} is nevertheless 
preserved. Indeed (\ref{wi+-}) implies for ${\rm i}\rho\to n\in\mathbb{N}_+$
\begin{equation}
 W_x\{I_{-{\rm i}\rho}(x),I_{{\rm i}\rho}(x)\}\sim \frac{2(-1)^n ({\rm i}\rho-n)}{x} \to 0.
\nonumber %%%\label{wj+-l}
\end{equation}
At the same time, the residues of $\Gamma(1-{\rm i}\rho)$ in the ${\rm i}\rho$ 
variable at those points are:
\begin{equation}
\left. \Res \Gamma(1-{\rm i}\rho)\right|_{{\rm i}\rho=n} =\frac{(-1)^{n}}{(n-1)!}\cdot
\label{gmres}
\end{equation}
Therefore, in the limit ${\rm i}\rho\to n\in\mathbb{N}_+$ ($k\to -k_n$),
\begin{eqnarray}
\lefteqn{
W_r\{f_+,f_-\} =
\Gamma(1-{\rm i}\rho)\Gamma(1+{\rm i}\rho) W_r \{I_{-{\rm i}\rho}(x),I_{{\rm i}\rho}(x)\}
}
\nonumber\\
&&
=\frac{2 n}{x} \left(\alpha {\rm e}^{-r/(2a)} \right)' =
\left. - \frac{n}{a}= - 2{\rm i}k\right|_{k=-k_n}.
\nonumber %%\label{wf+-l}
\end{eqnarray}
Analogously for ${\rm i}\rho\to -n$, or $k\to k_n$.

Following the present analysis and that of Ref. \cite{AM2h},
the redundant poles (zeros) correspond to the points where the irregular solution $f_-(k,r)$ 
($f_+(k,r)$) and the Jost function ${\cal F}_-(k)$ (${\cal F}_+(k)$) become {\em singular}
in the upper (lower) half of the complex $k$-plane. The origin of those singularities 
is that in analytic continuation 
of a parameter of two linearly {\em independent} 
solutions ({\rm i.e.} $\{I_{{\rm i}\rho}(z),I_{-{\rm i}\rho}(z)\}$ in the present case)
one cannot exclude that one ends up with linearly {\em dependent} 
solutions at a discrete set (which can be infinite) of isolated points in the parameter complex plane
({\rm i.e.} ${\rm i}\rho\to n\in\mathbb{Z}$ in the present case).
In view of the {\em factorization} (\ref{fctrz}) of each $f_\pm(k,r)$,
the above singularities of $f_\pm$ and ${\cal F}_\pm$ are 
in fact indispensable for preserving the fundamental identity 
$W\{f_+,f_-\}=-2{\rm i}k$ \cite[eq. (12.27)]{RN}. Without the above singular
behaviour of $f_\pm$ and ${\cal F}_\pm$ one would in fact face discontinuities 
of $W\{f_+,f_-\}=-2{\rm i}k$ for any ${\rm i}\rho\to\mathbb{Z}$.
The above singular behaviour is also essential in preserving 
the classical statement that if $f_+$ and $f_-$ exist, 
they are linearly {\em independent}, except when $k=0$ \cite[p. 336]{RN}, {\em i.e.} 
at the point where $W\{f_+,f_-\}=-2{\rm i}k\to 0$. Without the above singular behaviour,
$f_+$ and $f_-$ would exist and be linearly {\em dependent} for any $k=\pm k_n\ne 0$.

\section{Heisenberg condition}
\label{sc:hgc}
%%%%%%%%%%%%%%%%%%%%%%%%%%%%%%%%%%%%%%%%%
The {\em completeness} relation involving continuous and 
discrete spectrum yields \cite[eq. (12.128)]{RN} 
\begin{equation}
\frac{2}{\pi} \int_0^\infty \frac{\phi^*(k,r)\phi(k,r')}{|{\cal F}_+(k)|^2}\,k^2 dk
+
\sum_l \frac{\phi_l^*(r)\phi_l(r')}{N_l^2} =\delta_{r,r'},
\label{hu11}
\end{equation}
where $\delta_{r,r'}=\delta(r-r')$ and $N_l^2=\int_0^\infty [\varphi_l(r)]^2\, dr$ 
is an $l$th bound state (if any) normalization constant.
In the limit $r\to\infty$ one gets from \cite[eqs. (12.35), (12.71), (12.73)]{RN} 
for $k>0$
\begin{equation}
\phi(k,r)\sim \frac{{\cal F}_+(k) {\rm e}^{{\rm i}\delta(k)}}{k}\, \sin[kr+\delta(k)],
\label{rscs}
\end{equation}
where $\delta(k)$ is the scattering phase-shift \cite[eq. (12.95)]{RN}.
Given that $S(k)={\rm e}^{2{\rm i}\delta(k)}$ for $k\in\mathbb{R}$,
one can, on using asymptotic form (\ref{rscs}) of 
regular solutions for $r,r'\gg 1$ in the completeness relation (\ref{hu11}), 
arrive at \cite[eq. (6)]{Ma2}, \cite[eq. (1.2)]{BPS}, \cite[eq. (13)]{Hu}
\begin{equation}
\int_{-\infty}^\infty S(k) {\rm e}^{{\rm i}k(r+r')}\, dk = \sum_l |C_l|^2 {\rm e}^{-|k_l|(r+r')},
\label{hgc}
\end{equation}
where $|C_l|^2$ can be determined from the asymptotic of $\phi_l$ divided by $N_l^2$ \cite{AM2h}.
Under the condition that the integral over the real axis can be closed by an infinite
semicircle $\gamma$ in the upper half $k$-plane, {\em i.e.}
\begin{equation}
\oint_\gamma S(k) {\rm e}^{{\rm i}k(r+r')}\, dk = 0,
\label{hu16}
\end{equation}
one arrives at the correspondence between the poles of the S-matrix and bound states,
\begin{equation}
\oint_{k=k_l} S(k)\, dk = |C_l|^2 >0,
\label{hu15}
\end{equation}
where $\oint_{k=k_l}$ stands for integration along a contour encircling a single isolated bound state.
This correspondence is known as the {\em Heisenberg condition} \cite{Ma2,tH,BPS}.

Because $C_l\ne 0$ only for physical bound states, one has $\oint_{k=k_l} S(k)\, dk \equiv 0$
in the present case. To this end we determine the overall contribution of redundant poles 
to the integral on the lhs of (\ref{hgc}) as the sum over all residues.
On making use of eqs. (\ref{ipmm}) and (\ref{gmres}) in (\ref{MaSrf}), one finds
the following residue in the ${\rm i}\rho$ variable for any $k_n$ (${\rm i}\rho=-n$),
\begin{eqnarray}
\Res  S(k_n) &=& \frac{I_{n}(\alpha) }{I_{n}(\alpha) \Gamma(n+1)}
\left(\frac{\alpha}{2}\right)^{2n}\, \Res  \Gamma(-n+1)
\nonumber\\
&=& \frac{1}{n!} \left(\frac{\alpha}{2}\right)^{2n}\, \frac{(-1)^{n}}{(n-1)!}
\nonumber\\
&=& \frac{(-1)^n }{n!(n-1)!} \left(\frac{\alpha}{2}\right)^{2n}.
\nonumber %%\label{MaSrf}
\end{eqnarray}
When converting from ${\rm i}\rho$ to $k$ as independent variable, 
the left hand side of (\ref{hu15}) for $n$-th redundant pole has alternating sign and {\em not} always 
yields a positive number (the latter being typical for the true bound states),
\begin{equation}
2\pi {\rm i}\, \Res_k S(k_n) 
 = \frac{\pi}{a} \frac{(-1)^n}{n!(n-1)!} \left(\frac{\alpha}{2}\right)^{2n}.
\label{rprs}
\end{equation}
The overall contribution of the redundant poles to the integral 
on the lhs of (\ref{hgc}) is
\begin{eqnarray}
\lefteqn{
2\pi {\rm i} \sum_{n=1}^\infty \Res_k S(k_n) {\rm e}^{-n(r+r')/(2a)}
}
\nonumber\\
&& 
=
\frac{\pi}{a} \sum_{n=1}^\infty \frac{(-1)^n}{n!(n-1)!} 
 \left[\frac{\alpha}{2}\, {\rm e}^{-(r+r')/(4a)}\right]^{2n}
\nonumber\\
&& 
= - \frac{\pi}{a}\,q J_1(q),
\label{tcrp}
\end{eqnarray}
where $q=(\alpha/2) {\rm e}^{-(r+r')/(4a)}$ \cite[eq. (9.1.10)]{AS}, \cite[eq. (10.2.2)]{Ol}.
The overall contribution is an oscillating function of $q$.

\section{Residues of virtual states}
\label{sc:vst}
%%%%%%%%%%%%%%%%%%%%%%%%%%%%%%%%%%%%%%%%%
At any root of $I_{\nu}(\alpha)$ one has \cite[(9.6.42)]{AS}, \cite[(10.38.1)]{Ol},
\begin{equation}
\frac{\partial I_{\nu}(\alpha)}{\partial \nu}=-\left(\frac{\alpha}{2}\right)^{\nu}
\sum_{m=0}^\infty \frac{\psi(m+1+\nu)}{\Gamma(m+1+\nu)} \frac{(\alpha/2)^{2m}}{m!},
\nonumber %%\label{Iress}
\end{equation}
with $\psi$ here being the digamma function \cite[(6.3.1)]{AS}.
Because $I_\nu(z)$ is an entire function of its order \cite{Ann},
the residue of the pole term $1/I_{-{\rm i}\rho}(\alpha)$ in (\ref{MaSrf}) 
on the negative imaginary axis in the complex $k$-plane can be obtained as
\begin{equation}
\Res_k \frac{1}{I_{-{\rm i}\rho}(\alpha)} =\frac{{\rm i}}{2a}
          \left. \frac{1}{\partial_\nu I_\nu (\alpha)}\right|_{\nu=-{\rm i}\rho}.
\label{Ires}
\end{equation}

On the other hand, at any root of $J_\nu(z)$ \cite[(9.1.64)]{AS}, \cite[(10.15.1)]{Ol},
\begin{equation}
\frac{\partial J_{\nu}(\alpha)}{\partial \nu}=- \left(\frac{\alpha}{2}\right)^{\nu}
\sum_{m=0}^\infty (-1)^m \frac{\psi(\nu+m+1)}{\Gamma(\nu+m+1)} \frac{(\alpha/2)^{2m}}{m!}\cdot
\nonumber %%\label{Jress}
\end{equation}
The residue of the pole term $1/J_{-{\rm i}\rho}(\alpha)$ in the attractive case is then
determined by (\ref{Ires}) with $I_{-{\rm i}\rho}$ substituted by $J_{-{\rm i}\rho}$.

\section{Discussion}
\label{sc:dis}
%%%%%%%%%%%%%%%%%%%%%%%
Although there are many exactly solvable models known, it is quite rare
that one can determine analytic structure of the S-matrix up to the slightest detail, 
including position of all its poles and their residues, such as in the present case of an
exponentially decaying potential. The latter can be regarded as an example 
of exactly solvable S-matrix model. There is a number of important lessons to be learned 
from the repulsive exponentially decaying potential example (\ref{edrp}) studied here 
and its attractive version dealt with in ref. \cite{AM2h}.

\subsection{Repulsive vs attractive exponentially decaying potential}
\label{sc:rvsa}
%%%%%%%%%%%%%%%%%%%%%%%%%%%%%%%%%%%%%%%%%%%%%%%%%%%%%%%%%%%%%%%%%%%%%%%%%%%%%%%
Formally, the repulsive case differs from the attractive one in that $\alpha\to i\alpha$.
According to the connection formula \cite[(9.6.3)]{AS}, \cite[(10.27.6)]{Ol},
\begin{equation}
J_\nu(iz)=J_\nu\left(z{\rm e}^{\pi{\rm i}/2} \right) \to {\rm e}^{\nu\pi{\rm i}/2} I_\nu(z).
\nonumber
\end{equation}
Therefore, with a hindsight, it is not surprising that the expressions for irregular solutions $f_\pm(k,r)$
(\ref{f+}), the Jost function, ${\cal F}_\pm$ (\ref{Fkd}), and the $s$-wave S-matrix (\ref{MaSrf})
in the repulsive case can be transformed into those in the attractive case \cite{AM2h} by substituting 
$I_{\mp \nu}$ for $J_{\mp \nu}$. Nevertheless, as it is clear from derivations, it was not obvious in advance
which particular form the resulting expressions would assume.

\subsection{How to distinguish between the redundant poles and true bound 
states}
\label{sc:dng}
%%%%%%%%%%%%%%%%%%%%%%%%%%%%%%%%%%%%%%%%%%%%%%%%%%%%%%%%%%%%%%%%%%%%%%%%%%%%%%%
On the physical sheet one can clearly distinguish between the redundant poles and true bound 
states at the level of the Jost functions: (i) redundant poles are the 
{\em singularities} of ${\cal F}_-(k)$,
whereas (i) true bound states are the {\em zeros} of ${\cal F}_+(k)$.
Any difference between the respective poles gets blurred only at the level of the S-matrix 
when the ratio $S(k)={\cal F}_-(k)/{\cal F}_+(k)$ is formed.

\subsection{Basis of solutions at the points ${\rm i}\rho\in\mathbb{Z}$}
\label{sc:nsmps}
%%%%%%%%%%%%%%%%%%%%%%%%%%%%%%%%%%%%%%%%%%%%%%%%%%%%%%%%%%%%%%%%%%%%%%%%%%%%%%%
One witnesses in the literature a surprising unabated inertia in selecting the respective 
pairs $\{J_{{\rm i}\rho}(x),J_{-{\rm i}\rho}(x)\}$ and
$\{I_{{\rm i}\rho}(x),I_{-{\rm i}\rho}(x)\}$ as the basis
of linearly {\em independent} solutions of eq. (\ref{dbfe}) in exponentially 
attractive and repulsive cases, 
in spite that each of them collapses into linearly {\em dependent} solutions 
for any ${\rm i}\rho\in\mathbb{Z}$ [cf. eq. (\ref{wi+-})]. 
Those choices goes back to the classical contributions of Ma \cite{Ma1,Ma2,Ma3} and stretch,
for instance, to recent treatments of 
(i) one-dimensional exponential potentials $V(x)$ on $x\in(-\infty,\infty)$ \cite{AGK} and 
(ii) scattering and bound states in scalar and vector exponential potentials 
in the Klein-Gordon equation \cite[Sec. 4]{CCC}. 
The collapse of $\{I_{{\rm i}\rho}(x),I_{-{\rm i}\rho}(x)\}$
into linearly dependent solutions for ${\rm i}\rho \pi\ne n\pi$, $n\in\mathbb{Z}$
then inevitably prompts false conclusion that $\varphi(r)\equiv 0$ at the integer values of ${\rm i}\rho$ 
\cite{Ma1,Ma2,AGK,CCC}. Indeed, on expressing $K_{{\rm i}\rho}(x)$ in terms of 
$I_{\pm {\rm i}\rho}(x)$ according to (\ref{imnuknum}), and, on substituting into (\ref{gedprsi}), 
the regular solution becomes
\begin{eqnarray}
\varphi(r) &=& \frac{\pi C}{2 \sin ({\rm i}\rho \pi)}\,
\left[ I_{-{\rm i}\rho} (\alpha) I_{{\rm i}\rho}\left(x \right) \right.
\nonumber\\
&& \left. -\, I_{{\rm i}\rho}(\alpha) \, I_{-{\rm i}\rho}\left(x\right)\right].
\label{gedprsir}
\end{eqnarray}
One notices immediately that the square bracket in (\ref{gedprsir})
vanishes identically for any ${\rm i}\rho\in\mathbb{Z}$.
However, because of the singular prefactor, it is obviously not true
that $\varphi(r)\equiv 0$ for ${\rm i}\rho\in\mathbb{Z}$: one recovers (\ref{gedprsi}) 
in the limit ${\rm i}\rho\to n\in\mathbb{Z}$. 

The vanishing of the square bracket in (\ref{gedprsi}) necessitates
to work with the basis $\{I_{{\rm i}\rho}(x),K_{{\rm i}\rho}(x)\}$. The latter basis never 
degenerate into linearly dependent solutions and is standard choice when treating electromagnetic scattering 
from dielectric objects \cite{BH,AMap}. Although one can arrive at (\ref{MaSrf}) independently in the basis 
$\{I_{{\rm i}\rho}(x),I_{-{\rm i}\rho}(x)\}$ for ${\rm i}\rho \pi\ne n\pi$, $n\in\mathbb{Z}$, 
the introduction of the Jost functions ${\cal F}_\pm (k)$ becomes necessary in order to 
define the S-matrix for ${\rm i}\rho\in\mathbb{Z}$.

\subsection{Heisenberg condition}
\label{sc:hsnb}
%%%%%%%%%%%%%%%%%%%%%%%%%%%%%%%%%%
The S-matrix (\ref{MaSrf}) is formed essentially by the ratio of two entire functions 
$I_{{\rm i}\rho}(\alpha)/\Gamma(1-{\rm i}\rho)$ and 
$I_{-{\rm i}\rho}(\alpha)/\Gamma(1+{\rm i}\rho)$ \cite{Ann,Gmf}.
Given that the class of potentials for which the S-matrix is analytic for $k=\infty$ is 
very limited \cite{Ps}, the S-matrix (\ref{MaSrf}) is expected to have 
an {\em essential singularity}  there. The essential singularity
appears more complicated than ${\rm e}^{-2{\rm i}kR}$ for potentials 
of finite range $R$ (i.e. $V(r)\equiv 0$ for $r>R$),
which is featuring in the infinite product representation of their
S-matrix \cite[eq. (16)]{vanK}, \cite[eq. (12.107a)]{RN}, \cite[eq. (42)]{Hu}.
The essential singularity ${\rm e}^{-2{\rm i}kR}$ for finite-range potentials 
can be seen as a consequence of that the contour integral (\ref{hu16}) vanishes 
\cite[Sec. 12.1.4]{RN}, \cite[Appendix]{Hu}. Because one does not expect (\ref{hu16}) 
to hold in the present infinite-range potential case, the essential singularity of 
the S-matrix (\ref{MaSrf}) cannot be of the type ${\rm e}^{-2{\rm i}kR}$. 
Given that the limits $R\to \infty$ and $k=\infty$ do not commute \cite[p. 363]{RN}, 
it is impossible to determine the essential singularity in our case as a limiting case of 
finite-range potentials. Note in passing that for any finite-range potential
the Jost functions ${\cal F}_\pm (k)$ can have 
only a {\em finite} number of zeros on the imaginary axis \cite[pp. 361-2]{RN}, whereas 
e.g. ${\cal F}_+ (k)$ in our infinite-range potential case has {\em infinite} 
number of zeros yielding redundant poles $k_n\to{\rm i}\infty$ of 
the S-matrix (\ref{MaSrf}) on the physical sheet.

Like in the attractive case \cite{AM2h}, 
the overall, in general non-zero, contribution, of redundant poles (\ref{tcrp}) implies that 
the equality in (\ref{hgc}) cannot be preserved if one had attempted to perform the integral 
on the lhs of (\ref{hgc}) by closing the integration over the real axis in (\ref{hgc}) 
by an infinite semicircle $\gamma$ in the upper half $k$-plane 
and replaced it by the sum of residues of all enclosed poles.
As a consequence, the integral over 
the real axis in (\ref{hgc}) cannot be probably closed by an infinite
semicircle $\gamma$ in the upper half $k$-plane. If it could be somehow closed,
one cannot exclude that a contribution of the contour integral (\ref{hu16}) will
cancel the contribution of (\ref{tcrp}) of redundant poles, thereby restoring
the asymptotic completeness relation (\ref{hgc}).
Another valid point is that the use 
of asymptotic form (\ref{rscs}) of 
regular solutions in the completeness relation (\ref{hu11}) imply that
the relation (\ref{hgc}) is not a rigorous identity. It involves only leading asymptotic
terms of regular solutions for $r,\, r'\to\infty$ leaving behind subleading terms,
which may also contribute exponentially small terms in (\ref{hgc}).

Unfortunately, surprising absence of exact results for Bessel functions 
of general complex order \cite{AS,Ol} provides a true obstacle in full analytic analysis of that issue. 
We can only determine that on selecting different sequences when approaching $k\to\infty$ 
along the positive imaginary axis one arrives at different limits. 
For instance, for any $k_n$ one has 
$S(k_n)=\infty$, and the latter applies obviously also to the limit on the sequence $k_n\to\infty$.
On the other hand, one finds that the S-matrix (\ref{MaSrf}) has the following 
limit on the sequence $k={\rm i}(n+\tfrac{1}{2})/(2a)$ (${\rm i}\rho=-n-\tfrac{1}{2}$), $n\to\infty$ 
(see supplementary material)
\begin{eqnarray}
S(k) &\sim& \frac{2(2n)!!}{(2n-1)!! \sqrt{2\pi(2n+1)} },
\nonumber %%\label{MaSlnf}
\end{eqnarray}
which is the same as in the attractive case \cite{AM2h}.

\section{Conclusions}
\label{sc:con}
%%%%%%%%%%%%%%%%%%%%%%%%%%%%%%%%%%%%%%%%%%%%%%%%%%%%%%
A repulsive exponentially decaying potential (\ref{edrp}) provided us with a unique 
window of opportunity for a detailed study of analytic properties of the S-matrix 
for a potential of infinite range.
Its deep understanding was facilitated thanks to largely forgotten 
Coulomb's results \cite{Clb} on zeros of the modified Bessel function
$I_\nu(x)$. The resulting S-matrix (\ref{MaSrf}) 
was shown to exhibit unexpectedly rich behaviour hiding beautiful structures 
which were revealed by its domain coloring in fig. \ref{fgSmt}. 
Much the same can be said about an attractive exponentially decaying potential studied earlier 
in ref. \cite{AM2h}, which resulting S-matrix was exhibited in fig. \ref{fgSmta}.
Despite the innocuous Schr\"odinger equation (\ref{edper}), which does not show 
any peculiarity for $k=\pm k_n$, the S-matrix (\ref{MaSrf}) has always infinite number 
of redundant poles at any $k=k_n=in/(2a)$ on the physical sheet,
even in the absence of a single bound state. (In the attractive case
this happens for $\alpha<2$ \cite{AM2h}). On the second sheet of the complex 
energy plane, the S-matrix has (i) infinite number of poles corresponding to virtual states and 
a (ii) finite number of poles corresponding to complementary pairs of resonances 
and anti-resonances (those are missing in the attractive case).

The origin of redundant poles and zeros was confirmed to be related to peculiarities of 
analytic continuation of a parameter of two linearly independent analytic solutions
of the Schr\"odinger equation (\ref{edper}). We have obtained analytic expressions for the 
Jost functions and the residues of the S-matrix (\ref{MaSrf}) at the 
redundant poles [eq. (\ref{rprs})]. The overall contribution of redundant poles to the asymptotic 
completeness relation (\ref{hgc}), provided that the residue theorem can be applied, 
was determined to be an oscillating function [eq. (\ref{tcrp})]. 

Given that redundant poles and zeros occur already for such a simple model
is strong indication that they could be omnipresent for potentials of infinite range.
Currently one can immediately conclude that the appearance of poles of ${\cal F}_+(k)$, 
and of the S-matrix, at $k_n={\rm i}n/2a$ for positive integers $n$ is a general feature 
of potentials whose asymptotic tail is exponentially decaying like ${\rm e}^{-r/a}$ \cite{Ps}. 
This is because essential conclusions of our analysis will not change 
if the exact equalities involving $r$-dependence were replaced by asymptotic ones.
Redundant poles and more complicated essential singularity of $S(k)$ 
at infinity imply that any generalization of modal expansion of the scattered field 
for an infinite range potential will be much more involved than 
for the potentials of finite range \cite{GTV,CPS}.

Our results for the attractive case \cite{AM2h} can be readily applied for the analysis of 
the $s$-wave Klein-Gordon equation with exponential scalar and vector potentials \cite[Sec. 4]{CCC}. 
At the same time a proper understanding of analytic structure of the S-matrix
of essentially a textbook model will do no harm when attempting to generalize the presented 
results in the direction of non-Hermitian scattering Hamiltonians \cite{SBK}.
Last but not the least, we hope to stimulate search for further exactly solvable S-matrix models.

%\acknowledgments
\section*{Acknowledgments}	
%%%%%%%%%%%%%%%%%%%%%%%%%%
The work of AEM was supported by the Australian Research Council and UNSW Scientia Fellowship.

\newpage

\appendix

\section{Zeros of $J_\nu(x)$ for fixed nonzero $x\in\mathbb{R}$ 
considered as a function of $\nu$}
\label{Jzeros}
%%%%%%%%%%%%%%%%%%%%%%%%%%%%%%%%%%%%%%%%%%%%%%%%%%%%%%%%
Because of its importance and access difficulty, we find it expedient to
summarize Coulomb's work \cite{Clb} here. 
Coulomb's proof is, to a large extent, based on the Lommel integration 
formula \cite[\& 5$\cdot$11(13)]{Wat}, \cite[(1.13.2.5)]{PBM2},
\begin{eqnarray}
\lefteqn{
\int^t t^{-1} Z_\mu^{(1)}(tz)Z_\nu^{(2)}(tz)\, dt =
}
\nonumber\\
&&
- \frac{tz}{\mu^2-\nu^2}
\left[Z_{\mu+1}^{(1)}(tz) Z_\nu^{(2)}(tz) - Z_\mu^{(1)}(tz) Z_{\nu+1}^{(2)}(tz) 
\right]
\nonumber\\
&&
+
\frac{1}{\mu+\nu} Z_\mu^{(1)}(tz)Z_\nu^{(2)}(tz),
\label{lmldcf}
\end{eqnarray}
where $Z_\nu^{(1,2)}(z)$ are any two linear combinations of cylindrical Bessel functions, 
$t\in\mathbb{R}$, $z\in\mathbb{C}$.

For fixed $z$ ($\ne 0$) each branch of $J_\nu(z)$ is entire in {\em complex} variable $\nu$.
For $x\in\mathbb{R}$ one has $\overline{J_\nu(x)}=J_{\bar\nu}(x)$ (cf. \cite[(9.1.10)]{AS}, \cite[(10.2.2)]{Ol}), 
which is the Schwarz reflection in $\nu$-variable.
Hence a complex zero $\nu=\nu_0$ of $J_\nu(z)$ implies that $\bar\nu_0$ is also zero, 
i.e. complex zeros occur 
necessarily in complex conjugate pairs. In what follows it is sufficient
to limit oneself to positive $x$, because, according to analytic continuation formula, 
$J_\nu(z {\rm e}^{m\pi i})={\rm e}^{m\nu \pi i} J_\nu(z)$,
$m\in\mathbb{Z}$ \cite[(9.1.35)]{AS}, \cite[(10.11.1)]{Ol}.

\subsection{Re $\nu>0$}
%%%%%%%%%%%%%%%%%%%%%%%%
According to (\ref{lmldcf}),
\begin{eqnarray}
\int_0^1 t^{-1} J_{\nu_0}(tx)J_{\bar\nu_0}(tx)\, dt =
\int_0^1 t^{-1} |J_{\nu_0}(tx)|^2\, dt = 0.
\label{lmldcfj}
\end{eqnarray}
The rhs of (\ref{lmldcf}) yields zero at the upper integration limit under the 
hypothesis that $J_{\nu_0}(x)=J_{\bar\nu_0}(x)=0$.
It is zero at the lower integration limit for Re $\nu_0>0$ in virtue of the asymptotic behaviour of
each of $J_{\nu_0}$ and $J_{\nu_0+1}$ in the limit $z\to 0$ \cite[(9.1.7)]{AS}, \cite[(10.7.3)]{Ol},
\begin{equation}
J_\nu(z)\sim \frac{z^\nu}{2^\nu \Gamma(\nu+1)}\qquad (\nu\ne -1, -2, -3, \ldots).
\nonumber %%\label{as9.1.7}
\end{equation}
Because the integrand in (\ref{lmldcfj}) is a positive real quantity, we have a contradiction:
there is no complex zero $\nu_0$ of $J_\nu(x)$ for $x\in\mathbb{R}$ and Re $\nu>0$ \cite[item 3.1]{Clb}.

\subsection{Re $\nu<0$}
%%%%%%%%%%%%%%%%%%%%%%%%
The Lommel integration formula (\ref{lmldcf}) can be used to prove that 
$J_\nu(x)$ with $x\in\mathbb{R}$ has no complex zero $\nu_0$ also for any Re $\nu_0<0$ \cite[item 3.2]{Clb}.
To this end one performs integral 
\begin{equation}
\int_1^\infty t^{-1} J_{\nu_0}(tx)J_{\bar\nu_0}(tx)\, dt = \int_1^\infty t^{-1} |J_{\nu_0}(tx)|^2\, dt.
\nonumber %%\label{lmldcfj1}
\end{equation}
This time the hypothesis $J_{\nu_0}(x)=J_{\bar\nu_0}(x)=0$ implies that the rhs of (\ref{lmldcf}) is 
zero at the lower integration limit.
At the upper integration limit one makes use of the asymptotic 
behaviour for $z\to \infty$ \cite[(9.2.1)]{AS}, \cite[(10.7.8)]{Ol},
\begin{equation}
J_\nu(z)\sim \sqrt{2/(\pi z)}\, \cos \left(z-\tfrac{1}{2}\,\nu\pi-\tfrac{1}{4}\, \pi\right)~~ (|\arg z|< \pi).
\nonumber %%\label{as9.2.1}
\end{equation}
The latter implies that only the square bracket on the rhs of the Lommel integration formula (\ref{lmldcf})
contributes at the upper integration limit,
\begin{eqnarray}
\lefteqn{
\int_1^\infty t^{-1} |J_{\nu_0}(tx)|^2\, dt
=-\frac{2}{\pi}\frac{1}{\nu_0^2-\bar\nu_0^2}\, \times
}
\nonumber\\
&&
 \left[
\cos \left(\omega -\tfrac{1}{2}\,\pi - \tfrac{i}{2}\,\mbox{Im }\nu_0\pi\right)
 \cos \left(\omega + \tfrac{i}{2}\,\mbox{Im }\nu_0\pi\right)
\right.
\nonumber\\
&&
\left.
-
\cos \left(\omega - \tfrac{i}{2}\,\mbox{Im }\nu_0\pi\right)
 \cos \left(\omega -\tfrac{1}{2}\,\pi + \tfrac{i}{2}\,\mbox{Im }\nu_0\pi\right)
\right],
\nonumber %%\label{lmldcfj1fp}
\end{eqnarray}
where $\omega=z-\tfrac{1}{2}\,\mbox{Re }\nu_0\pi-\tfrac{1}{4}\, \pi$.
Because the square bracket can be recast as
\begin{eqnarray}
\lefteqn{
\sin \left(\omega - \tfrac{i}{2}\,\mbox{Im }\nu_0\pi\right)\cos \left(\omega + \tfrac{i}{2}\,\mbox{Im }\nu_0\pi\right)
}
\nonumber\\
&&
-
\cos \left(\omega - \tfrac{i}{2}\,\mbox{Im }\nu_0\pi\right)\sin \left(\omega + \tfrac{i}{2}\,\mbox{Im }\nu_0\pi\right)
\nonumber\\
&&
= - \sin (i\, \mbox{Im }\nu_0\pi)=-i\sinh (\mbox{Im }\nu_0\pi),
\nonumber %%\label{cospp}
\end{eqnarray}
one obtains eventually 
\begin{equation}
\int_1^\infty t^{-1} |J_{\nu_0}(tx)|^2\, dt
=\frac{2i}{\pi}\frac{1}{\nu_0^2-\bar\nu_0^2}\, \sinh (\mbox{Im }\nu_0\pi).
\label{lmldcfj1f}
\end{equation}
Then the lhs of (\ref{lmldcfj1f}) is positive, whereas its rhs
\begin{eqnarray}
\lefteqn{
\frac{2i}{\pi}\frac{1}{(\nu_0-\bar\nu_0)(\nu_0+\bar\nu_0)}\, \sinh (\mbox{Im }\nu_0\pi)=
}
\nonumber\\
&&
\frac{1}{\pi(\nu_0+\bar\nu_0)}\, \frac{\sinh (\mbox{Im }\nu_0\pi)}{\mbox{Im }\nu_0}<0,
\nonumber %%\label{lmldcfj1fr}
\end{eqnarray}
when $\nu_0+\bar\nu_0=2\mbox{Re }\nu_0<0$. Hence
$J_\nu(x)$ with $x\in\mathbb{R}$ has no complex zero for $\mbox{Re }\nu<0$ \cite[item 3.2]{Clb}.

\subsection{Re $\nu=0$}
%%%%%%%%%%%%%%%%%%%%%%%%
Eventually, one can prove that $J_\nu(x)$ cannot have purely imaginary zeros \cite[item 3.3]{Clb}.
If it were some, then obviously $\bar\nu_0=-\nu_0$. However the pair $\{J_{\nu_0}(x), J_{-\nu_0}(x)\}$
provides a basis of linearly independent solutions of the Bessel equation for any non-integer $\nu_0$.
Because its Wronskian cannot vanish, it is impossible 
that $J_{\nu_0}(x)= J_{-\nu_0}(x)=0$ for some value of $x$.

In order to prove the simplicity of zeros of $J_\nu(x)$, Coulomb \cite[item 3.4]{Clb} made use of
Watson's formula \cite[\&13.73(2)]{Wat}
\begin{eqnarray}
\lefteqn{
J_\nu(x) \frac{\partial Y_\nu(x)}{\partial\nu}-Y_\nu(x) \frac{\partial J_\nu(x)}{\partial\nu}
}
\nonumber\\
&&
= -\frac{4}{\pi}\, \int_0^\infty K_0(2x\sinh t) {\rm e}^{-2\nu t}\, dt<0,
\nonumber %%\label{w13.73.2}
\end{eqnarray}
where the inequality is valid for $\nu\in\mathbb{R}$, $x>0$. 
(Note in passing that $K_\nu(x)$ is real and positive for real order $\nu\ge -1$ and $x>0$ \cite[(9.6.1), (9.6.6)]{AS}.)
Obviously, $J_\nu(x)$ and $\partial J_\nu(x)/\partial\nu$ cannot vanish simultaneously.

\section{Zeros of $I_\nu(x)$ for fixed nonzero $x\in\mathbb{R}$ 
considered as a function of $\nu$}
\label{Izeros}
%%%%%%%%%%%%%%%%%%%%%%%%%%%%%%%%%%%%%%%%%%%%%%%%%%%%%%%%
For fixed $z$ ($\ne 0$) each branch of $I_\nu(z)$ is entire in $\nu$.
For $x\in\mathbb{R}$ one has $\overline{I_\nu(x)}=I_{\bar\nu}(x)$ (cf. \cite[(9.6.10)]{AS}, \cite[(10.25.2)]{Ol}), 
which is the Schwarz reflection in $\nu$-variable.
Hence a complex zero $\nu=\nu_0$ implies that also $\bar\nu_0$ is zero, i.e. complex zeros occur 
necessarily in complex conjugate pairs.
In virtue of analytic continuation formula $I_\nu(z {\rm e}^{m\pi i})={\rm e}^{m\nu \pi i} I_\nu(z)$,
$m\in\mathbb{Z}$, \cite[(9.6.30)]{AS}, \cite[(10.34.1)]{Ol}, it is again sufficient
to limit oneself to positive $x$.

Coulomb ingenious proof on the impossibility of complex ({\em i.e.} with nonzero imaginary part) 
zeros of $I_\nu(x)$ for Re $\nu =\nu_1 >-3/2$ \cite[item 4.2]{Clb} is 
based on the generalized Neumann's formula \cite[\&13.72(2)]{Wat},
\begin{eqnarray}
\lefteqn{
I_\mu(x) I_\nu(x) =
}
\nonumber\\
&&
 \frac{2}{\pi}\, \int_0^{\pi/2} I_{\mu+\nu}(x)(2x\cos\theta) \cos[(\mu-\nu) \theta]\, d\theta,
\label{w13.72.2}
\end{eqnarray}
which is valid for Re $(\mu+\nu)>-1$. 

First, for a complex conjugate pair of $\mu$ and $\nu$ with Re $\nu =\nu_1 >-1/2$ one obtains from (\ref{w13.72.2})
\begin{eqnarray}
|I_{\nu}(x)|^2 &= &
 \frac{2}{\pi}\, \int_0^{\pi/2} I_{2 \nu_1}(x)(2x\cos\theta) \cosh (2 \nu_2 \theta)\, d\theta
 \nonumber\\
&>& I_{\nu_1}^2(x)>0,
\nonumber %%\label{clb4.2}
\end{eqnarray}
where Im $\nu=i\nu_2$. Note in passing that $I_\nu(x)$ is real and positive 
for real order $\nu\ge -1$ and $x>0$ \cite[(9.6.1), (9.6.6)]{AS}.
This proves that there are no complex zeros of $I_\nu(x)$ for Re $\nu >-1/2$.

On returning back to (\ref{w13.72.2}) in the special case when $\mu\ne \nu$, yet Im $\nu=$ - Im $\mu$ and Re $\mu >-1/2$,
\begin{eqnarray}
\lefteqn{
I_\mu(x) I_\nu(x) =
}
\nonumber\\
&& \frac{2}{\pi}\, \int_0^{\pi/2} I_{\mu_1+\nu_1}(x)(2x\cos\theta)
 \cos[(\mu-\nu) \theta]\, \, d\theta,
\label{w13.72.2s}
\end{eqnarray}
where 
\begin{eqnarray}
\lefteqn{
\cos[(\mu-\nu) \theta] = \cos[(\mu_1-\nu_1) \theta]\sinh[(\mu_2-\nu_2) \theta]
}
\nonumber\\
&& 
- i \sin[(\mu_1-\nu_1) \theta]\cosh[(\mu_2-\nu_2) \theta].
\nonumber %%\label{imcos}
\end{eqnarray}
For $\mu_1=\nu_1$ the imaginary part of $\cos[(\mu-\nu) \theta]$ is zero but its real part is positive.
For $\mu_1\ne\nu_1$, the imaginary part of $\cos[(\mu-\nu) \theta]$ will maintain constant sign on the integration 
interval $(0,\pi/2)$ in (\ref{w13.72.2s}), and thus prevents it from vanishing, when $|\mu_1-\nu_1|<2$. 
Combined together with the hypothesis 
Re $\mu >-1/2$ that ensures $I_\mu(x)\ne 0$, and $\mu_1+\nu_1>-1$ necessary for the validity
of (\ref{w13.72.2}), the task is to find the smallest possible $\nu_1$ that would comply with all the above conditions.
This is obviously Re $\nu =-3/2+\epsilon$ with some infinitesimal $\epsilon$ (in which case $\mu_1=1/2+\epsilon'$, $\epsilon'<\epsilon$).
Therefore $I_\nu(x)$ does not have any complex zero for Re $\nu>-3/2$ \cite[item 4.2]{Clb}.

That $I_\nu(x)$ does not have any complex zero for Re $\nu\ge 0$ can be derived also independently 
by tweaking Coulomb's proof for $J_\nu(x)$ \cite[items 3.1,3]{Clb}.
Indeed, one can arrive at a Lommel integration formula also for the modified Bessel functions,
\begin{eqnarray}
\lefteqn{
\int^t t^{-1} Z_\mu^{(1)}(tz)Z_\nu^{(2)}(tz)\, dt =
}
\nonumber\\
&&
\frac{tz}{\mu^2-\nu^2}
\left[Z_{\mu+1}^{(1)}(tz) Z_\nu^{(2)}(tz) - Z_\mu^{(1)}(tz) Z_{\nu+1}^{(2)}(tz) 
\right]
\nonumber\\
&&
+
\frac{1}{\mu+\nu} Z_\mu^{(1)}(tz)Z_\nu^{(2)}(tz),
\label{lmbldcf}
\end{eqnarray}
where $Z_\nu^{(1,2)}(z)$ are any two linear combinations of {\em modified} cylindrical Bessel functions, 
$t\in\mathbb{R}$, $z\in\mathbb{C}$. 
The Lommel integration formula (\ref{lmbldcf}) differs from (\ref{lmldcf}) merely in the opposite sign
in front of the square bracket on the rhs. 
Formula (\ref{lmbldcf}) can be verified by differentiating both sides
and using the defining modified Bessel equation \cite[(9.6.1)]{AS}, \cite[(10.25.1)]{Ol}.
One can thus readily repeat the arguments that has led to (\ref{lmldcfj}),
and thereby exclude the complex zeros of $I_{\nu}(x)$ for Re $\nu>0$.

Any purely imaginary zero $\nu_0$ of $I_\nu(x)$ can be excluded by essentially the same argument as in \cite[item 3.3]{Clb}.
If it were some $\nu_0$, then obviously $\bar\nu_0=-\nu_0$. However the pair $\{I_{\nu}(x), I_{-\nu}(x)\}$
provides a basis of linearly independent solutions of the Bessel equation for any non-integer $\nu$.
Because its Wronskian (\ref{wi+-}) cannot vanish, it is impossible that $I_{\nu_0}(x)= I_{-\nu_0}(x)=0$.

$I_\nu(z)$ has the following asymptotic 
behaviour for $z\to \infty$ \cite[(9.7.1)]{AS}, \cite[(10.40.1)]{Ol},
\begin{equation}
I_\nu(z)\sim \frac{{\rm e}^z}{\sqrt{2 \pi z}}\left(1-\frac{4\nu^2-1}{8z}\right)\qquad (|\arg z|< \pi/2).
\nonumber %%\label{as9.7.1}
\end{equation}
Therefore, one cannot perform $\int_1^\infty$ as in eq. (\ref{lmldcfj1f}) in the case of $J_\nu(z)$.

\section{$S(k)$ for $k\to\infty$}
%%%%%%%%%%%%%%%%%%%%%%%%%%%%%%%%%
For $k={\rm i}(n+\tfrac{1}{2})/(2a)$ (${\rm i}\rho=-n-\tfrac{1}{2}$) the S-matrix (8) of the main text becomes
\begin{eqnarray}
S(k)= \frac{I_{-n-\tfrac{1}{2}}(\alpha) \Gamma(\tfrac{1}{2} -n)}{I_{n+\tfrac{1}{2}}(\alpha) \Gamma(\tfrac32+n)}
\left(\frac{\alpha}{2}\right)^{2n+1},
\nonumber %%\label{MaSlr}
\end{eqnarray}

According to (14) of the main text (cf. \cite[eq. (9.6.2)]{AS}, \cite[eq. (10.27.2)]{Ol}),
\begin{equation}
I_{-n-\tfrac{1}{2}}(z) =I_{n+\tfrac{1}{2}}(z) +(-1)^n \frac{2}{\pi}\, K_{n+\tfrac{1}{2}}(z).
\label{as9.1.2}
\end{equation}
According to \cite[eq. (10.41.1-2)]{Ol},
for positive real values of $\nu$ in the limit $\nu\to\infty$
\begin{eqnarray}
I_{\nu}(z) &=& \frac{1}{\sqrt{2\pi\nu}} \left(\frac{{\rm e}z}{2\nu}\right)^\nu, 
\nonumber\\
K_{\nu}(z) &=& \sqrt{\frac{\pi}{2\nu}} \left(\frac{{\rm e}z}{2\nu}\right)^{-\nu},
\nonumber 
\end{eqnarray}
which, on combining with (\ref{as9.1.2}), enables one to arrive at
\begin{equation}
\frac{I_{-n-\tfrac{1}{2}}(\alpha)}{I_{n+\tfrac{1}{2}}(\alpha)} 
 \sim 1+ 2 (-1)^n \left(\frac{2n+1}{{\rm e}\alpha}\right)^{2n+1}.
\nonumber
\end{equation}

At the same time on repeating the defining relation $\Gamma(z+1)=z\Gamma(z)$ one has
\begin{eqnarray}
\Gamma(n+\tfrac32) &=& \frac{(2n+1)!!}{2^{n+1}}\Gamma(\tfrac{1}{2}),
\nonumber\\
\Gamma(\tfrac{1}{2}) &=& (-1)^n \frac{(2n-1)!!}{2^n}\Gamma(\tfrac{1}{2}-n),
\nonumber\\
\frac{\Gamma(\tfrac{1}{2} -n)}{\Gamma(\tfrac32+n)} &=& (-1)^n \frac{2^{2n+1}}{(2n-1)!! (2n+1)!!}\cdot
\nonumber
\end{eqnarray}

Therefore, the S-matrix (8) of the main text has the following 
limit for $k={\rm i}(n+\tfrac{1}{2})/(2a)$ (${\rm i}\rho=-n-\tfrac{1}{2}$), $n\to\infty$,
\begin{eqnarray}
S(k) &\sim& \frac{2}{(2n-1)!!(2n+1)!!} \left(\frac{2n+1}{{\rm e}}\right)^{2n+1}
\nonumber\\
&\sim& \frac{2(2n+1)!}{(2n-1)!!(2n+1)!! \sqrt{2\pi(2n+1)} }
\nonumber\\
&\sim& \frac{2(2n)!!}{(2n-1)!! \sqrt{2\pi(2n+1)} },
\nonumber %%\label{MaSlnf}
\end{eqnarray}
where Stirling's formula has been used to arrive at the $2$nd line.
The resulting asymptotic behaviour is the same as in the attractive case \cite{AM2h}.

%%%%%%%%%%%%%%%%%%%%%%%%%%


\begin{thebibliography}{99}
\bibitem{Gmf}
$\Gamma(z)$ has no zeros, so the reciprocal gamma function 
$1/\Gamma(z)$ is an entire function \cite{AS}.
\end{thebibliography}
\end{document}